\documentclass[12pt]{iopart}

\usepackage{amssymb}
\usepackage{color,ulem}
\usepackage{graphicx}
\usepackage{dcolumn}
\usepackage{bm}
\usepackage{color}
\newcommand{\nn}{\nonumber}
\newcommand{\beq}{\begin{eqnarray}}
\newcommand{\eeq}{\end{eqnarray}}

\begin{document}

\title[Topological phases of quasi-one-dimensional fermionic atoms]
{Topological phases of quasi-one-dimensional fermionic atoms with
a synthetic gauge field}

\author{Takeshi Mizushima}
\address
{Department of Physics, Okayama University, Okayama 700-8530, Japan,}
\address
{Department of Physics and Astronomy, Northwestern University, Evanston,
Illinois 60208, USA} 
\author[label3]{Masatoshi Sato}
\address
{Department of Applied Physics, Nagoya University, 464-8603, Japan}

\begin{abstract}

We theoretically investigate the effect of intertube tunneling in
topological superfluid phases of a quasi-one-dimensional Fermi gas with
a Rashba-type spin-orbit interaction. It is shown that the effective
Hamiltonian is analogous to that of a nanowire topological
superconductor with multibands.
Using a hidden mirror symmetry in the system, we introduce a new
 topological number that ensures the existence of non-Abelian Majorana
 zero modes even in the presence of intertube tunneling.  
It is demonstrated
 from the full numerical calculation of self-consistent equations that
 some of Majorana modes survive against the intertube tunneling, when
 the number of one-dimensional tubes is odd in the $y$-direction.  
We also discuss a generalization of our consideration to nanowire
 topological superconductors.
\end{abstract}

%\begin{keyword}
%Topological Superconductivity \sep Cold Atoms \sep Edge States \sep
%Majorana Fermions %
%\end{keyword}

%\end{frontmatter}

% \linenumbers

%---------------------------------------- Sec. Introduction ----------------------------------------
\section{Introduction}
\label{introduction}

Majorana fermions are real fermions which are equivalent to their own
anti-particles. Since the pioneering works by 
Read and Green~\cite{read} and Kitaev~\cite{kitaev2001}, search for
their elusive fermions opened 
an exciting new chapter in condensed matter physics. Original works in
Refs.~\cite{read} and \cite{kitaev2001} 
predicted that the mysterious
fermions exist as zero-energy quasiparticles bound at vortices and edges
of a spinless $p$-wave superconductor. Subsequently, tremendous progress
has succeeded in extending the platform for realizing Majorana fermions
to some categories of so-called {\it topological
superconductors}~\cite{review1,review2,schnyder,kitaev}. The remarkable
consequence of the self-charge conjugate property of Majorana fermion is
non-Abelian braiding statistics, where a pair of Majorana zero modes are
created or annihilated by braiding their host
vortices~\cite{ivanov}. Hence, Majorana fermions possessing non-Abelian
braiding statistics can provide a promising platform for fault-torelant
topological quantum computation~\cite{nayak}. Moreover, it has recently
been unveiled that zero-energy quasiparticles exhibit multifaceted
properties, not only as a Majorana fermion but also as odd-frequency
Cooper pair correlation~\cite{tanaka1,tanaka2,yokoyama,tanuma,higashitani,daino,tsutsumi,asano}. 

An ideal candidate for realizing non-Abelian Majorana zero modes was a
chiral $p$-wave superconductor with half-quantum vortices, where the
low-lying quasiparticles are effectively spinless~\cite{ivanov}. A
half-quantum fluxoid has been observed in mesoscopic annular rings of
Sr$_2$RuO$_4$~\cite{budakian}, while the half-quantum vortices in
superconductors are energetically unstable against integer vortices
because of the absence of screening mechanism of spin
current~\cite{chung}. The another candidate of a chiral $p$-wave
superfluid is the A-phase of superfluid $^3$He confined to a restricted
geometry with sub-micron thickness~\cite{tsutsumi2010}. The
thermodynamic stability of half-quantum vortices in such a superfluid is
not trivial, because the Fermi liquid corrections which favor vortices
with spin flow rather than mass flow are competitive to the strong
coupling correction due to the spin fluctuation which stabilizes integer
vortices without spin flow~\cite{kawakami}.  

On the contrary, it was demonstrated that a conventional $s$-wave superconductor can harbor non-Abelian Majorana zero modes~\cite{sato2003,fu,sato2009v2,sato2010v2,czhang}. The key finding lies in the two-dimensional Rashba-type spin-orbit interaction in background normal fermions, where the non-Abelian anyons are due to the phase twist of the spin-orbit interaction. The electron bands split by the spin-orbit interaction effectively convert the $s$-wave pairing to $p\pm ip$ pairing. A strong Zeeman field drives the quantum phase transition from a non-topological phase without non-Abelian anyons to a topological phase. This finding provides an another approach to the realization of topological quantum computation in condensed matters. Indeed, it has been proposed that topological superconductivity and Majorana fermions can be realized in a one-dimensional (1D) semiconducting wire proximity-coupled with an $s$-wave superconductor~\cite{sau,lutchyn,alicea,oreg,brouwer,alicea2011}, where semiconductors, such as InSb, have very large $g$-factor and strong spin-orbit interaction. The signature of Majorana fermions has recently been observed through zero bias conductance peaks in a nanowire topological superconductor~\cite{mourik,deng,das} and unconventional Josephson effect in hybrid superconductor-topological insulator devices~\cite{williams}. 

Apart from superconducting materials, cold atoms with a $p$-wave
Fershbach resonance~\cite{TM2008,radzihovsky} or with a synthetic gauge
field~\cite{sato2009v2,sato2010v2} offer an alternative playground for
Majorana fermions. A spin-orbit coupling with equal Rashba and
Dresselhaus strengths can be synthetically induced by applying Raman
lasers to atomic gases with hyperfine spin degrees of
freedom~\cite{nist}, whose practical scheme was first pointed out by Liu {\it et al.}~\cite{liu}.
This scheme has recently been implemented using fermionic $^6$Li~\cite{mit} and $^{40}$K atoms~\cite{china}. 
In addition, schemes for creating
Rashba and Dresselhaus spin-orbit coupling and three-dimensional
analogue to Rashba spin-orbit coupling have theoretically been
proposed~\cite{juzeliunas,campbell,anderson}. A 1D geometry with 
a resonantly interacting Fermi gas can be
implemented by using a two-dimensional optical lattice, which was
already utilized to search the Fulde-Ferrell-Larkin-Ovchinnikov (FFLO)
state in a spin-imbalanced Fermi gas~\cite{rice}. 
Hence, cold atoms with
Raman laser-induced spin-orbit coupling provide not only a promising
platform for realizing Majorana
fermions~\cite{sato2009v2, Liu2012,Liu2012v2,Wei,Zhai} but also an ideal
system to study competition of various exotic superfluid phases including
topological and FFLO phases. Furthermore, it has been proposed that 
topologically non-trivial phases, such as topological insulating phases, can be realized in a 
Fermi gas with a non-Abelian and Abelian gauge field in an optical lattice~\cite{goldman,bermudez,mazza,goldman09,bermudez10,ruostekoski,javanainen,ruostekoski08}.

In this paper, we study topology and quasiparticle spectra of a quasi-1D
Fermi gas with a Rashba-type spin-orbit coupling. It has been
demonstrated in Refs.~\cite{Liu2012,Liu2012v2,Wei} that a pure 1D Fermi
gas with spin-orbit coupling is accompanied by exactly zero-energy
states bound at the end points of atomic clouds. However, since an
actual experiment is performed on a bundle of weakly coupled tubes,
intertube tunneling effects are not negligible. We here demonstrate that
the intertube coupling plays an important role on determining the
topological properties of Fermi gases and the effective Hamiltonian is
analogous to nanowire topological superconductors with
multibands~\cite{Potter,lutchyn2011,stanescu,kells,Tewari}. 
The existence of non-Abelian Majorana zero modes is ensured by introducing 
a new topological number associated with a mirror symmetry.  

This paper is organized as follows. We begin in Sec. 2 by introducing a
tight-binding model for a bundle of 1D Fermi gases with spin-orbit
coupling. In Sec. 3, we clarify the topology of such a system, where a
one-dimensional winding number is protected by a hidden mirror
symmetry. It turns out that this system provides a cold atom analogue to
nanowire topological superconductors with multibands. In Sec. 4, based
on fully numerical calculations of self-consistent equations, we study
the intertube tunneling effect on quasiparticle spectra. The final
section is devoted to conclusions and discussions. 
Throughout this paper, we set $\hbar\!=\! k_{\rm B} \!=\! 1$ and the
repeated Greek indices imply the sum over $x, y, z$.

%\section{Overview on synthetic gauge fields}

%---------------------------------------- Sec. Model ----------------------------------------
\section{Array of one-dimensional tubes}
\label{model}

We here start with the Hamiltonian for spin-orbit coupled two-component fermionic atoms with an $s$-wave attractive interaction, $g$,
\beq
\mathcal{H} = \int d{\bm r} {\bm \Psi}^{\dag}({\bm r})\left[
\epsilon ({\bm r}) + \mathcal{S} ({\bm r})
\right]{\bm \Psi}({\bm r})
+ g\int d{\bm r} \psi^{\dag}_{\uparrow}({\bm r})\psi^{\dag}_{\downarrow}({\bm r})
\psi _{\downarrow}({\bm r})\psi _{\uparrow}({\bm r}),
\label{eq:H}
\eeq
where ${\bm \Psi} \!\equiv\! [\psi _{\uparrow}, \psi_{\downarrow}]^{\rm T}$ denotes the fermionic field operators with up- and down-spins. The single-particle Hamiltonian density is defined as
$\epsilon ({\bm r}) 
\!=\! -\frac{1}{2m}{\bm \nabla}^2  - \mu _{\rm cp} + V_{\rm pot}({\bm r}) 
- h_{\mu}\sigma _{\mu}$ with a confinement potential $V_{\rm pot}$ and $\sigma _{\mu}$ being the Pauli matrices in spin space. The
Zeeman field ${\bm h}$ is naturally induced by
implementing the spin-orbit coupling through two-photon Raman
process~\cite{nist}. In Eq.~(\ref{eq:H}), $\mathcal{S}$ describes
the spin-orbit coupling term, which is expressed in general as 
\beq
\mathcal{S}({\bm r}) = iA_{\mu \nu}\sigma _{\nu}\partial _{\mu}.
\eeq

Within the mean-field approximation, the Hamiltonian in Eq.~(\ref{eq:H}) can be diagonalized in terms of the quasiparticle states. The quasiparticle states with the energy $E_{n}$ are obtained by solving the so-called Bogoliubov-de Gennes (BdG) equation~\cite{TMJPSJ}, 
\beq
\mathcal{H}({\bm r}){\bm \varphi}_n({\bm r}) = E_n {\bm \varphi}_n({\bm r}), 
\eeq
where ${\bm \varphi} \!=\! [u_{n,\uparrow},u_{n,\downarrow},v_{n,\uparrow},v_{n,\downarrow}]^{\rm T}$ denotes the wavefunction of quasiparticles, where $u_{n,\sigma}$ and $v_{n,\sigma}$ describe the wave functions of the particle- and hole-components, respectively. The BdG Hamiltonian density is given as
\beq
\mathcal{H}({\bm r}) = \left(
\begin{array}{cc}
\epsilon ({\bm r}) + \mathcal{S} ({\bm r}) & i\sigma _y \Delta ({\bm r}) \\
-i\sigma _y \Delta ({\bm r}) & - \epsilon^{\ast}({\bm r}) - \mathcal{S}^{\ast}({\bm r})
\end{array}
\right),
\label{eq:Hfull}
\eeq
where $\Delta ({\bm r})$ is an $s$-wave pair potential with a contact interaction $g$, defined as $\Delta ({\bm r}) \!=\! g\langle \psi _{\downarrow}({\bm r})\psi _{\uparrow}({\bm r})\rangle$.
Note that the BdG Hamiltonian density in Eq.~(\ref{eq:Hfull}) holds the particle-hole symmetry, $\tau_{x}\mathcal{H}({\bm r})\tau_x \!=\! - \mathcal{H}^{\ast}({\bm r})$, where $\tau_{\mu}$ denotes the Pauli matrices in particle-hole space.

\begin{figure}
\centering
\includegraphics[width=80mm]{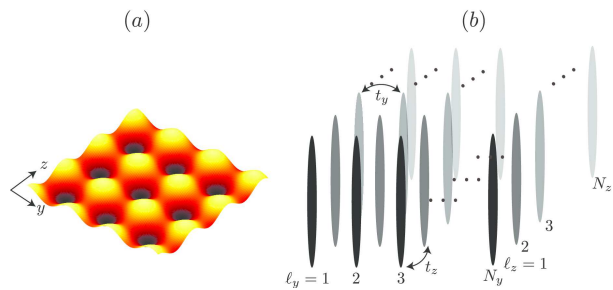}
\caption{(a) Two-dimensional optical lattice potential and (b) schematic picture of the calculated system.}
\label{fig:pot}
\end{figure}

In order to isolate Majorana zero modes from the higher energy
quasiparticle states, the fermionic atoms are confined by a
two-dimensional optical lattice in the $y$-$z$ plane in addition to a
shallow harmonic potential along the $x$-direction, as shown in
Fig.~\ref{fig:pot}. The system under this confinement potential is
regarded as a two-dimensional array of $N_y \! \times \! N_z$ 1D
tubes. The quasiparticle wavefunction is expanded in terms of the
functions $f_{\ell_y}(y)$ and $f_{\ell_z}(z)$ localized at $(x,y_{\ell_y}, z_{\ell_z})$ as ${\bm
\varphi} ({\bm r}) = \sum _{{\bm \ell}} {\bm
\varphi}_{\bm \ell}(x)f_{\ell_y}(y)f_{\ell_z}(z)$ with ${\bm
\ell}=(\ell_y, \ell_z)$, where $\ell_y \!=\!1, \cdots, N_y$ and $\ell_z \!=\! 1, \cdots, N_z$. 
Employing the tight-banding
approximation in the $y$-$z$ plane, the Hamiltonian in
Eq.~(\ref{eq:Hfull}) reduces to  
\beq 
\mathcal{H}^{\rm eff}(x) = \left(
\begin{array}{cc}
\underline{\epsilon}(x) + \underline{\mathcal{S}}(x) & i\sigma _y \underline{\Delta}(x) \\
-i\sigma _y \underline{\Delta}^{\dag} (x) & - \underline{\epsilon^{\ast}}(x) - \underline{\mathcal{S}}^{\ast}(x)
\end{array}
\right),
\label{eq:Heff}
\eeq
where the pair potential $\Delta ({\bm r})$ is transformed to a $2N_yN_z\times 2N_yN_z$ matrix for ${\bm \ell}$ and ${\bm \ell}^{\prime}$, $\underline{\Delta}(x)$. The single-particle Hamiltonian density $\underline{\epsilon}(x)$ is a $2N_yN_z\times 2N_yN_z$ matrix given by
\beq
[\underline{\epsilon} (x)]_{{\bm \ell},{\bm \ell}^{\prime}}  &=& 
\left(
- \frac{1}{2m}\frac{d^2}{dx^2} 
- \mu _{\rm cp} + V_{{\bm \ell}}(x) - h_{\mu}\sigma _{\mu}
\right)\delta _{{\bm \ell},{\bm \ell}^{\prime}} \nn \\
&& - t_y (\delta _{{\bm \ell},{\bm \ell}^{\prime}+ \hat{\bm e}_y}
+\delta _{{\bm \ell},{\bm \ell}^{\prime}- \hat{\bm e}_y})
- t_z (\delta _{{\bm \ell},{\bm \ell}^{\prime}+ \hat{\bm e}_z}
+ \delta _{{\bm \ell},{\bm \ell}^{\prime}- \hat{\bm e}_z}).
\eeq
The effective potential $V_{\bm \ell}(x)$ is given as $V_{\bm \ell}(x)
\!\equiv\! \frac{1}{2}m\omega^2_x x^2 + V_{\bm \ell}$. 
Here we set ${\bm \ell} \!=\! \ell_y \hat{\bm e}_y +\ell_z \hat{\bm e}_z$ 
with $\hat{\bm e}_y \!=\! (1, 0)$ and $\hat{\bm e}_z \!=\! (0,1)$. 
The hopping energies between intertubes are denoted by
$t_y$ and $t_z$. The spin-orbit coupling term also reduces to
\begin{eqnarray}
[\underline{\mathcal{S}} (x)]_{{\bm \ell},{\bm \ell}^{\prime}}  = i \sigma _{\mu}\left[
A_{x\mu} \partial _x \delta _{{\bm \ell},{\bm \ell}^{\prime}}
+ \tilde{A}_{y\mu}(\delta _{{\bm \ell},{\bm \ell}^{\prime}+ \hat{\bm e}_y}
-\delta _{{\bm \ell},{\bm \ell}^{\prime}- \hat{\bm e}_y})
\right.
\nonumber\\
\left.
+ \tilde{A}_{z\mu}(\delta _{{\bm \ell},{\bm \ell}^{\prime}+ \hat{\bm
e}_z}
- \tilde{A}_{z\mu}\delta _{{\bm \ell},{\bm \ell}^{\prime}- \hat{\bm e}_z})\right], 
\end{eqnarray}
where $\tilde{A}_{\mu\nu}$ describes an effective non-Abelian gauge field, 
$\tilde{A}_{y\mu} \equiv \int f^{\ast}_{\ell_z}(z) f^{\ast}_{\ell_y+1}(y) A_{y\mu} f_{\ell_y}(y)f_{\ell_z}(z)dydz
$. 
Note that a quasi-1D Fermi gas has been reported in Ref.~\cite{rice}
with a two-dimensional optical lattice potential. 
The number of tubes is typically about 
$N_y \!\times\! N_z \! \sim \mathcal{O}(10\times 10)$, so the system
should be treated as a finite system. 

Under the tight-binding approximation, the resulting BdG equation reduces to an effectively 1D equation along the $x$-axis, 
\beq
\mathcal{H}^{\rm eff}_{{\bm \ell},{\bm \ell}^{\prime}}(x) {\bm \varphi}_{n,{\bm \ell}^{\prime}}(x) 
= E_n {\bm \varphi}_{n,{\bm \ell}}(x),
\label{eq:bdg}
\eeq
which is numerically solved with the finite element method implemented with the discrete variable representation~\cite{TM2010v2}.
The BdG equation (\ref{eq:bdg}) is selfconsistently coupled with the gap equation for the pair potential, $[\underline{\Delta} (x)]_{{\bm \ell},{\bm \ell}^{\prime}}  \!=\! \Delta _{\bm \ell} (x) \delta _{{\bm \ell},{\bm \ell}^{\prime}}$, 
\beq
\Delta _{\bm \ell}(x) 
= g \sum _{E_n}\left[
u_{n,\uparrow, {\bm \ell}}(x)v^{\ast}_{n,\downarrow, {\bm \ell}}(x) f(E_n) 
+ u_{n,\downarrow, {\bm \ell}}(x)v^{\ast}_{n,\uparrow, {\bm \ell}}(x) f(-E_n)
\right],
\label{eq:gap}
\eeq
where $f(E) \!=\! 1/(e^{E/T}+1)$ is the Fermi distribution function at a temperature $T$. In addition, the chemical potential $\mu$ is determined so as to preserve the total particle number
\beq
N = \sum _{E_n, \sigma}\int d{\bm r}\left[
\left| u_{n,\sigma, {\bm \ell}}(x)\right|^2f(E_n) + \left| v_{n,\sigma, {\bm \ell}}(x)\right|^2f(-E_n)
\right].
\label{eq:number}
\eeq
The sum in Eqs.~(\ref{eq:gap}) and (\ref{eq:number}) is taken over $E_n \!\in\! [0,E_{\rm c}]$, where $E_{\rm c}$ denotes the energy cutoff. The effective 1D coupling constant, $g$, in Eq.~(\ref{eq:gap}) is expressed in terms of an effective 1D scattering length $a_{\rm 1D}$ as~\cite{liu2007}
\beq
g = - \frac{2}{ma_{\rm 1D}} = - \frac{2}{m}\frac{a_{\rm 3D}/d_{\perp}}{Aa_{\rm 3D}-d_{\perp}},
\label{eq:g}
\eeq
where $a_{\rm 3D}$ denotes a three-dimensional scattering length and $d_{\perp}$ is the characteristic harmonic oscillator length in $y$- and $z$-axes. The constant $A$ is given as $A \!\approx\! 1.0326$. From Eq.~(\ref{eq:g}), the 1D scattering length $a_{\rm 1D}$ is expressed in terms of $a_{\rm 3D}$ as $a_{\rm 1D}\!=\! d_{\perp}(A- d_{\perp}/a_{\rm 3D})$. In the definition of $a_{\rm 1D}$, the sign is opposite to that of $a_{\rm 3D}$. Hence, the Cooper pairing state in a 1D Fermi gas can be stabilized by a positive $a_{\rm 1D}$ which corresponds to $a_{\rm 3D} < 0$ realized in $^6$Li atoms. It is also seen from Eq.~(\ref{eq:g}) that the pairing interaction, $g$, can be controlled by changing the characteristic length scale of the confinement potential in the $y$-$z$ plane, $d_{\perp}$.

Before closing this section, we mention the validity of the mean-field theory in a quasi-1D Fermi gas. For a pure 1D system, in general, the quantum fluctuation plays a critical role, which violates the long-range ordering. Hence, the mean-field approximation employed in the current work might not work very well in a pure 1D system. Based on the direct comparison of the mean-field theory with the Bethe ansatz solutions, however, Liu {\it et al.} \cite{liu2007} demonstrated that the mean-field theory provides a useful description in weakly or moderately interacting regimes of a pure 1D Fermi gas. Furthermore, the quantum fluctuation is suppressed by introducing the intertube tunneling $t_y$ and $t_z$, which involves the crossover of the single-particle dispersion from 1D to 3D. 

%---------------------------------------- Sec. Topology of the effective Hamiltonian ----------------------------------------
\section{Topology of the effective Hamiltonian}

We here consider a two-dimensional Rashba-type spin-orbit interaction, 
\beq
A_{\mu\nu}\sigma _{\nu}\partial _{\mu} = \kappa _{x} \sigma _y \partial _x - \kappa _y \sigma _x \partial _y.
\eeq
To clarify the topological property of the effective Hamiltonian, we
here ignore the shallow trap potential along $x$-direction, i.e.,
$\omega _x \!\rightarrow\! 0$. Then, the effective Hamiltonian in
Eq.~(\ref{eq:Heff}) is rewritten with $-i\partial _x \!\rightarrow\! k$
as 
\beq
\mathcal{H}^{\rm eff}_{{\bm \ell}, {\bm \ell}'}(k) 
&=& \left[ 
\epsilon^{(0)}_{{\bm \ell},{\bm \ell}^{\prime}}(k) 
- \left\{ h_x\sigma _x +h_z \sigma _z +\kappa _x k \sigma
_y \right\}
\delta _{{\bm \ell},{\bm \ell}^{\prime}}\right]\tau _z \nn \\
&& - \Delta _{{\bm \ell},{\bm \ell}^{\prime}}\sigma _y \tau _y 
+ \left[  h_y\sigma_y \delta _{{\bm \ell},{\bm \ell}^{\prime}} + i \tilde{\kappa}_y\sigma _x 
\left\{\delta _{{\bm \ell},{\bm \ell}^{\prime}+ \hat{\bm e}_y}
-\delta _{{\bm \ell},{\bm \ell}^{\prime}- \hat{\bm e}_y}\right\} \right] \tau_0,
%-h_y \sigma _y .
\label{eq:Heffk}
\eeq
where $\epsilon^{(0)}_{{\bm \ell},{\bm \ell}^{\prime}}$ describes the
single-particle Hamiltonian density without the Zeeman term. %and $\tau
%_{\mu}$ denotes the Pauli matrices in particle-hole space. 
%The additional term $\mathcal{H}_1$ is composed of the contributions
%from the spin-orbit coupling and Zeeman field, 
%\beq
%\mathcal{H}_1 = 
%\eeq
Here, without the loss of generality, $\Delta _{\bm \ell}$ is assumed to
be real. We have also assumed $h_y=0$.
It is seen from Eq.~(\ref{eq:Heffk}) that the array of 1D tubes
with a spin-orbit interaction is analogous to a
semiconductor-superconductor nanowire with $N$-th electron
bands~\cite{Tewari}, where $N  \!\equiv\! N_y \!\times\!N_z$. 

We find that our system supports two different topological numbers.
The first one is the 1D ${\mathbb Z}_2$ topological number for
class D topological phases:
Because of superfluidity, the BdG Hamiltonian (\ref{eq:Heffk}) has the
particle-hole symmetry, which allows us to define the 1D ${\mathbb
Z}_2$ topological number.
The 1D ${\mathbb Z}_2$ number is defined as
\begin{eqnarray}
\nu=\frac{1}{\pi}\int_{-\pi}^{\pi}dk A(k)+\mbox{mod. 2},
\label{eq:z2top}
\end{eqnarray}
where $A(k)$ is the geometrical phase, 
\begin{eqnarray}
A(k)=i\sum_{E_n(k)<0}\sum _{\bm \ell}
\langle {\bm \varphi}_{n,{\bm \ell}}(k)|\partial_k
 {\bm \varphi}_{n,{\bm \ell}}(k)\rangle,
\end{eqnarray} 
with $|{\bm \varphi}_{n,{\bm \ell}}(k)\rangle$ the Bloch wave function of an negative energy state
of ${\cal H}^{\rm eff}(k)$. 
When $\nu$ is odd (even), the system is topologically non-trivial
(trivial). 

The second topological number comes from a remnant of a mirror
reflection symmetry of the system.
If one temporary neglects the Zeeman fields ${\bm h}$, our system is
invariant under the mirror reflection to the $zx$-plane, as well as the
time-reversal.
Once the Zeeman fields are applied, the mirror symmetry is lost, but a
combination of the mirror reflection and the time-reversal is still
preserved if $h_y=0$.
Consequently, the Hamiltonian $\mathcal{H}_{\rm eff}(k)$ with $h_y=0$ holds the
following ${\bm Z}_2$ symmetry,
\beq
\mathcal{T}\mathcal{M}_{zx}\mathcal{H}_{\rm eff}(k)
\mathcal{M}_{zx}^{\dagger}\mathcal{T}^{-1} = \mathcal{H}^{\ast}_{\rm eff}(-k),
\label{eq:z2}
\eeq
where $\mathcal{T} \!=\! i\sigma_y K$ is the time-reversal operator
with the complex conjugate operator $K$, and 
$\mathcal{M}_{zx}=i\sigma_yU$  the mirror reflection operator.  
Here $U$ is the operator flipping the $y$-component of ${\bm
\ell}=(\ell_y, \ell_z)$, 
\begin{eqnarray}
U_{{\bm \ell}, {\bm \ell}'}=\delta_{\ell_y, N_y+1-\ell_y'} 
\,\delta_{\ell_z,\ell_z'}.
\end{eqnarray}

Combining the ${\bm Z}_2$ symmetry with the particle-hole symmetry, 
$\mathcal{C}\mathcal{H}_{\rm eff}(k)\mathcal{C}^{-1} = - \mathcal{H}^{\ast}_{\rm eff}(-k)$, 
we define the chiral symmetry operator, 
$\Gamma \!=\! \mathcal{T}\mathcal{M}_{zx}\mathcal{C} \!=\! \tau _xU$, where $\mathcal{C} \!=\! \tau _x K$ denotes the particle-hole transformation operator with complex conjugation operator $K$. Then, it turns out that $\Gamma$ is anti-commutable with the effective Hamiltonian
\beq
\left\{ \Gamma, \mathcal{H}_{\rm eff} (k)\right\} = 0.
\eeq
This implies that the BdG Hamiltonian $\mathcal{H}_{\rm eff}(k)$ holds
the chiral symmetry.
%The BdG Hamiltonian $\mathcal{H}_{\rm eff}(k)$ can be off-diagonalized in a basis where the operator $\Gamma$ is diagonalized to $\mathcal{U}\Gamma\mathcal{U}^{\dag} \!=\! {\rm diag}({1_{N \!\times\!N}, -1_{N \!\times\!N}})$,
%\beq
%\mathcal{U}\mathcal{H}_{\rm eff}(k)\mathcal{U}^{\dag} = \left(
%\begin{array}{cc}
%0 & \mathcal{Q}(k) \\ \mathcal{Q}^{\dag}(k) & 0
%\end{array}
%\right),
%\eeq
%where $\mathcal{U}$ is a $4N$-dimensional unitary matrix and
%$\mathcal{Q}_{{\bm \ell},{\bm \ell}^{\prime}}(k) \!\equiv\!
%\epsilon^{(0)} _{{\bm \ell},{\bm \ell}^{\prime}} - ( h_x\sigma _x + h_z
%\sigma _z +  \kappa k \sigma _y) \delta _{{\bm \ell},{\bm
%\ell}^{\prime}} + i \Delta _{{\bm \ell},{\bm \ell}^{\prime}} \sigma
%_y$. 
Then, the 1D winding number is defined as~\cite{sato2009, sato2011,TM2012,wen}
\beq
w = -\frac{1}{4\pi i} \int^{\infty}_{-\infty} dk 
{\rm tr}\left[\Gamma{\cal H}^{-1}_{\rm eff}(k) \partial_k {\cal H}_{\rm
eff}(k)\right],
\label{eq:w}
\eeq
which takes an integer.
A similar 1D winding number was considered for 2D 
and pure 1D Rashba
superconductors,  
where $\Gamma$ in Eq.(\ref{eq:w}) is replaced by
$\tau_x$ \cite{sato2009, TS2012}.
The above expression (\ref{eq:w}) is a generalization of these cases into
multi-tube systems.

As a consequence of the bulk-edge correspondence, these two 1D
topological numbers ensure the existence of zero energy states
appearing in the end points of 1D tubes. 
Here we note that the parities of these two topological numbers coincide
with each other,
\begin{eqnarray}
(-1)^{\nu}=(-1)^w, 
\end{eqnarray}
which implies that $w$ can be nonzero even when $\nu$ is trivial, but
the opposite is not true.
Therefore, the actual number of the zero energy states is determined by $w$
unless the ${\bm Z}_2$ symmetry (\ref{eq:z2}) is broken macroscopically.
In addition, the particle-hole
symmetry of superfluidity results in the Majorana property of
the zero energy state, where the creation operator $\gamma^{\dag}_{E\!=\! 0}$ is
equivalent to its own annihilation, $\gamma _{E\!=\! 0} \!=\!
\gamma^{\dag}_{E\!=\! 0}$~\cite{sato2011,TM2012}. In summary, the
winding number $w$ ensures the
existence of Majorana zero modes, whereas once the ${\bm Z}_2$ symmetry
is broken by turning on $h_y$ for example, $w$ becomes
ill-defined and the 1D ${\mathbb Z}_2$ number $\nu$ in Eq.(\ref{eq:z2top})
determines the topological stability of the Majorana zero modes. 

In the case of a two-dimensional disk geometry, where the
two-dimensional optical lattice potential is absent, the topological
property is characterized by the first Chern
number~\cite{sato2010v2}. The nontrivial value of the Chern number
ensures the existence of the gapless state localized at the
circumference of the atomic cloud. In addition, the edge states carry
the macroscopic mass current.  
 
%------------------------------------ Sec. Intertube tunneling effect ------------------------------------
\section{Intertube tunneling effect}

Let us start with a pure 1D system with $t_y \!=\! t_z \!=\!
\tilde{\kappa}_y \!=\! 0 $. Here,
the BdG equation (\ref{eq:bdg}) coupled with the gap equation
(\ref{eq:gap}) is numerically solved with the set of parameters: $T
\!=\! 0$, $E_{\rm c} \!=\! 4E_{\rm F}$, and $\kappa _x \!=\! 1$. The Zeeman
field is applied along $\hat{\bm z}$-axis, which does not break the
${\bm Z}_2$ symmetry in Eq.(\ref{eq:z2}): ${\bm h} \!=\! (0,0,h)$. In realistic
situations~\cite{nist,Zhai}, the strength of the spin-orbit coupling,
$\kappa _x$, depends on the wave length of applied lasers. Throughout
this work, we fix the pairing interaction, $\gamma \!=\! 1.4$, where
$\gamma \!=\! \frac{1}{\pi\sqrt{N}}(\frac{d}{a_{\rm 1D}})$ is the
dimensionless coupling constant~\cite{liu2007}, where $d \!=\!
\sqrt{1/(m\omega _x)}$ is the harmonic oscillator length. The total particle
number is fixed to be $N \!=\! 200$ in each tube, where the Fermi energy
per one tube is $E_{\rm F} \!=\! 100\omega _x$. 

Before going to numerical results, 
we first clarify the effect of the mean-field potential 
associated with the on-site interaction $g$, which is dropped 
in the BdG Hamiltonian density in Eq.~(\ref{eq:Hfull}). 
The on-site interaction term changes the single-particle Hamiltonian density in Eq.~(\ref{eq:Hfull})
to ${\epsilon} ({\bm r}) \!\rightarrow\! {\epsilon} ({\bm r}) + g \rho ({\bm r}) + gm ({\bm r})\sigma _z$, where $\rho ({\bm r}) $ and $m({\bm r})$ denote the local particle and spin densities. 
Within the local density approximation,
the maximum value of the local density $\rho (x)$ in an ideal Fermi gas confined to a 1D harmonic oscillator is estimated as $\max\rho (x) \!\equiv\! \rho _0 \!=\! 2\sqrt{N}/\pi d$. 
The dimensionless parameter $\gamma$ gives rough estimation about 
the ratio of the local potential and the Fermi energy, $\gamma \sim g \rho _0/E_{\rm F}$, 
where in our calculated system, $\mu \!\sim\! E_{\rm F}$. 
Since the potential term changes the local chemical potential and the local Zeeman field,
it quantitatively alters quasiparticles with finite 
energies and the critical Zeeman field above which Majorana zero modes appear.
However, the on-site interaction term does not affect the topological properties 
associated with Majorana zero modes, where the mirror symmetry is preserved. 
Therefore, we here ignore the effect of the on-site interaction term in the BdG equation.

For the pure 1D system, topology of the BdG Hamiltonian for each tube is
characterized by the winding number $w$ in Eq.~(\ref{eq:w}) with $N_y=N_z=1$.
With spatially constant $\Delta$ and $\mu$, $w$ is given by 
\beq
w=\frac{1}{2\pi i}\int_{-\infty}^{\infty}dk \partial_k
\ln\left[\det {\cal Q}(k)\right],
\eeq
where $\mathcal{Q}(k) \!\equiv\!
\epsilon^{(0)}(k) - ( h_x\sigma _x + h_z
\sigma _z +  \kappa k \sigma _y)+ 
i \Delta\sigma_y$.
Then, it is found that $w
\!=\! 1$ when $|h| \!>\! h_{\rm c}
\!\equiv\! \sqrt{\mu^2 + \Delta^2}$. 
Therefore the lower magnetic field regime is
the topologically trivial phase and the critical field at $h_{\rm c}$
involves the topological phase transition.
For our system, however, a more careful consideration is needed.
Since fermionic atoms are confined by a trap potential, 
a spatially inhomogeneous superfluid is realized naturally. 
In addition, the inhomogeneous pair potential $\Delta(x)$, which is
self-consistently determined by the gap equation and the BdG equation,
depends on the Zeeman fields significantly.
In contrast to
semiconductor-superconductor junction systems, 
these two characteristics cannot be neglected.
This means that within the local density
approximation, the critical field $h_{\rm c}$ and 1D winding number $w$
should be replaced by $h_{\rm c}(x) \!\equiv\! \sqrt{\mu^2(x) + \Delta^2(x)}$
and $w(x)$, where $\mu(x)$ is the local chemical potential including the
confinement potential along the $x$-axis. 
The inhomogeneity and self-consistency of $\Delta (x)$ and $\mu (x)$ play a
critical role on the topological property. 

\begin{figure}
\centering
\includegraphics[width=70mm]{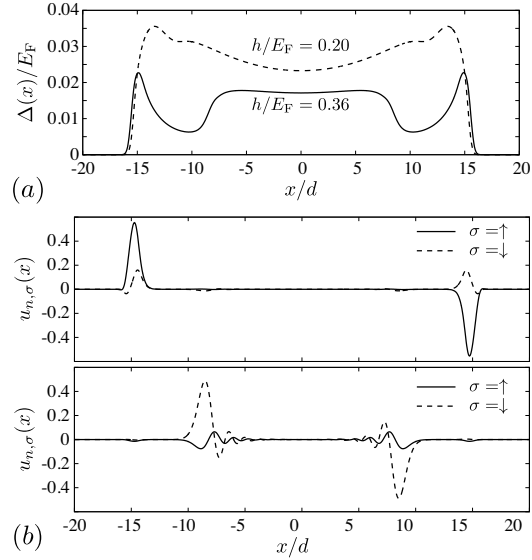}
\caption{(a) Spatial profiles of the pair potential $\Delta (x)$ in a
 pure 1D system with $t_y \!=\! t_z \!=\! 0$ and $h_y \!=\! 0$. (b)
 Wavefunctions of the lowest (upper panel) and second lowest (lower
 panel) energy eigenstates at $h \!=\! 0.36 E_{\rm F}$, where the
 energies are $E\!=\! 3.77\!\times\! 10^{-10}E_{\rm F}$ and
 $7.28\!\times\! 10^{-9}E_{\rm F}$. } 
\label{fig:gap}
\end{figure}

Figure~\ref{fig:gap}(a) shows
the spatial profile of $\Delta (x)$ at $h \!=\! 0.20E_{\rm F}$ and $0.36
E_{\rm F}$, where the harmonic trap potential along the $x$-axis,
$\frac{1}{2}m\omega^2_x x^2$, is taken into account. In the case of $h
\!=\! 0.36 E_{\rm F}$, the intermediate region between $8.8d \!\lesssim\! |x|
\!\lesssim \! 14.6d$, where the pair
potential has a dip, 
becomes topologically non-trivial, {\it i.e.} it
satisfies $h \!>\! h_{\rm c} (x)$ and $w(x) \!\neq\! 0$, while 
the inner region within $|x|\!\lesssim\! 8.8 d$ is not.
It is also found that the outermost region is not topological again,
 because the local chemical potential, $\mu(x) \!\equiv\! \mu -
\frac{1}{2}m\omega^2_xx^2$, changes its sign at the Thomas-Fermi edge $x
\!=\! x_{\rm TF} \!\equiv\! d\sqrt{2\mu/\omega _x} \!\sim\! 14 d$ and
its magnitude becomes large so as $h<h_{\rm c}(x)$.    

\begin{figure}
\centering
\includegraphics[width=90mm]{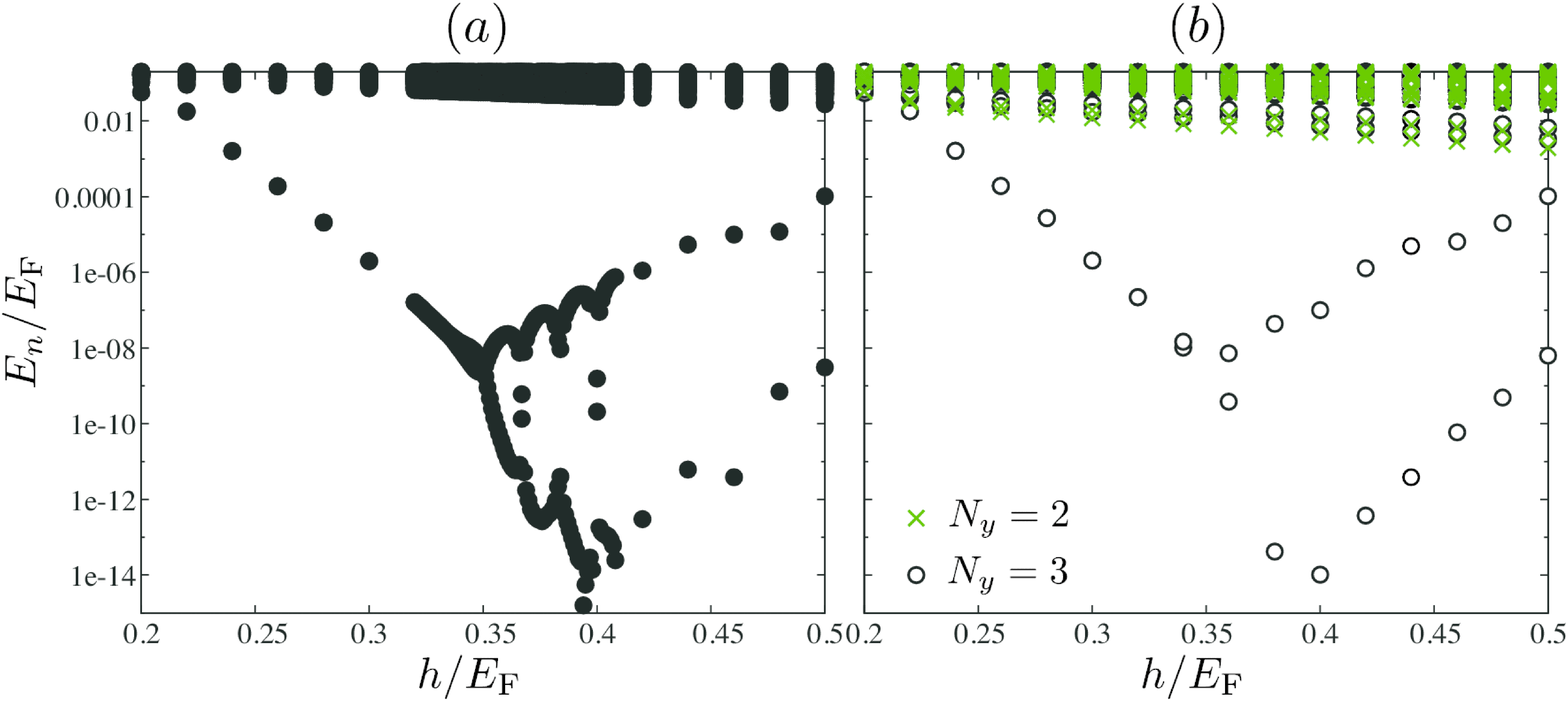} \\
\includegraphics[width=70mm]{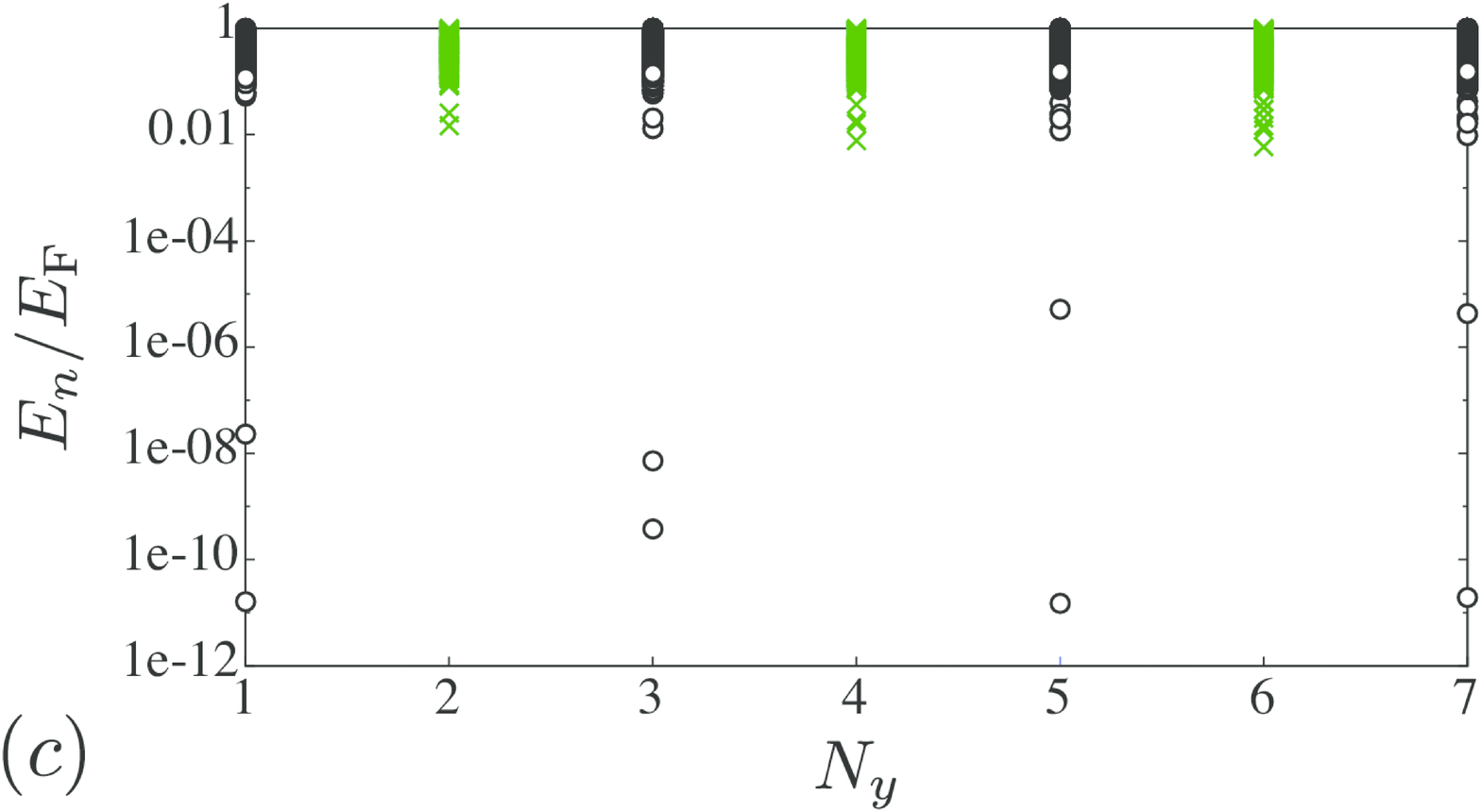}
\caption{Field-dependence of quasiparticle energy spectra: (a) A pure 1D
 case (closed circles) and (b) quasi-1D cases for $N_y \!=\! 2$ (green
 crosses) and $3$ (opened circles). (c) Quasiparticle spectra as a
 function of $N_y$ at $h \!=\! 0.36E_{\rm F}$ which corresponds to the
 topological phase in the case of $N_y \!=\! 1$. The green crosses and
 opened circles denote the energy spectra for even and odd $N_y$'s, respectively. In all the data, the hopping $t_y$ and the strength of the spin-orbit interaction $\tilde{\kappa}_y$ is set to be $t_y \!=\! 0.01E_{\rm F}$ and $\tilde{\kappa}_y/\kappa _x \!=\! 0.5$.}
\label{fig:egns}
\end{figure}

In the absence of the hopping between tubes, four zero modes appear in
each tube. At $h \!=\! 0.36 E_{\rm F}$, $|w|\!=\! 1$ can be realized in
the region within $8.8 d \!\lesssim\! |x| \!\lesssim\! 14.6 d$,
otherwise $w \!=\! 0$. As shown in Fig.~\ref{fig:gap}(b), the lowest and
second lowest energy states are tightly bound at the phase boundaries at
$x \!\sim\! \pm 8d$ and $\pm 15 d$.  
%It is also worth mentioning that the lowest (second lowest) energy state originates from the large (small) fermi surface. 
We summarize the field-dependence of the energy spectrum $E_n$ in
Fig.~\ref{fig:egns}(a). In our calculated system, the level spacing due to the harmonic potential along the $x$-axis
is given as $0.01 E_{\rm F}$. Thus, in Fig.~\ref{fig:egns}(a), the quasiparticles with $E\!>\! 0.01E_{\rm F}$ can be regarded as the ``{\it continuum} states'' and those with $E\!<\! 0.01E_{\rm F}$ is referred to as the bound states. The quasiparticle states having $E\!\ll\! 0.01E_{\rm F}$ can be referred to as the ``{\it zero-energy}'' states. 
It is seen from Fig.~\ref{fig:egns}(a) that the low-lying eigenenergies go
to {\it zero} as $h$ increases, because of the
interference between two zero modes localized at $x/d \!\sim\! \pm 8$
and $\pm 15$~\cite{TM2010v2,Cheng}. In the Zeeman field regime higher than $0.35E_{\rm F}$, the {\it zero energy} states are splited to two branches. It turns out that the upper (lower) branches correspond to the quasiparticle states bound at the inner (outer) edges at $x \!\sim\! \pm 8d$ ($\pm 15d$), as displayed in Fig.~\ref{fig:gap}(b). As $h$ further increases, the amplitude of the pair potential $\Delta (x)$ decreases. This implies that the wave function of the zero energy states spreads, giving rise to the hybridization of zero energy states.

Now, let us clarify how the intertube tunneling affect the low-lying
quasiparticle spectra. The bundle of one-dimensional tubes are coupled
with each other through the hoppings $t_y$ and $t_z$ and the spin-orbit
interactions $\tilde{\kappa}_y$. 
In this situation, 
the topological winding number $w$ is not defined for each tube, but is
defined only for a whole system of tubes.
%becomes ill-defined in such a
%situation, because the ${\bm Z}_2$ symmetry is explicitly broken by the
%term with $\tilde{\kappa}_y$. 
As a result, some of Majorana zero modes become non zero modes as is
shown below.
Note that if the hopping $t_z$ is small enough, it does not
change the winding number and the topological property because of the
two-dimensionality of the Rashba-type spin-orbit coupling. 
This implies
that zero energy states are dispersionless and insensitive to
$N_z$. Figure \ref{fig:egns}(b) shows the field-dependence of
low-lying quasiparticle energies for $N_y \!=\! 2$ and $3$, where we fix
the parameters, $t_y \!=\! 0.01E_{\rm F}$ and $\tilde{\kappa}_y/\kappa
_x \!=\! 0.5$. In the case of $N_y \!=\! 2$, the zero energy states are 
lifted to finite energies by their hybridization 
through the intertube tunneling $t_y \!\neq\! 0$. In contrast, as shown 
in Fig.~\ref{fig:egns}(b), the zero energy state in the case of $N_y \!=\! 3$ 
survives even in the presence of the finite intertube tunneling. 
Figure \ref{fig:egns}(c) summarizes the low-lying spectra as
a function of $N_y$, where $h$ is fixed to be $h/E_{\rm F} \!=\! 0.36$
which corresponds to the topological phase in the case of $N_y \!=\!
1$. It is clear that the zero energy states exist when $N_y$ is odd,
while in the case of even $N_y$'s, the intertube tunneling lifts the
low-lying states from the zero energy.  

\begin{figure}
\centering
\includegraphics[width=65mm]{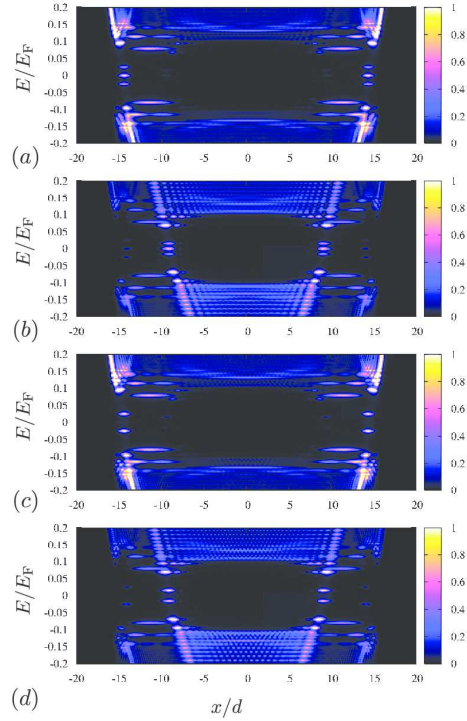}
\caption{Local density of states $\mathcal{N}_{\sigma,{\bm \ell}}(x,E)$
 defined in Eq.~(\ref{eq:ldos}) in the case of $N_y \!=\! 3$ and $h
 \!=\! 0.36E_{\rm F}$. The LDOS at $\ell_y \!=\! 1$ for $\sigma \!=\!
 \uparrow$ and $\downarrow$ is displayed in (a) and (b),
 respectively. The LDOS at $\ell_y \!=\! 3$ is same as that at $\ell _y \!=\! 1$, when $N_y \!=\! 3$. 
 (c) and (d) shows the LDOS for $\sigma \!=\! \uparrow$
 and $\downarrow$ at $\ell_y \!=\! 2$. The other parameters are same as
 those in Fig.~\ref{fig:egns}.} 
\label{fig:ldos}
\end{figure}

\begin{figure}
\centering
\includegraphics[width=65mm]{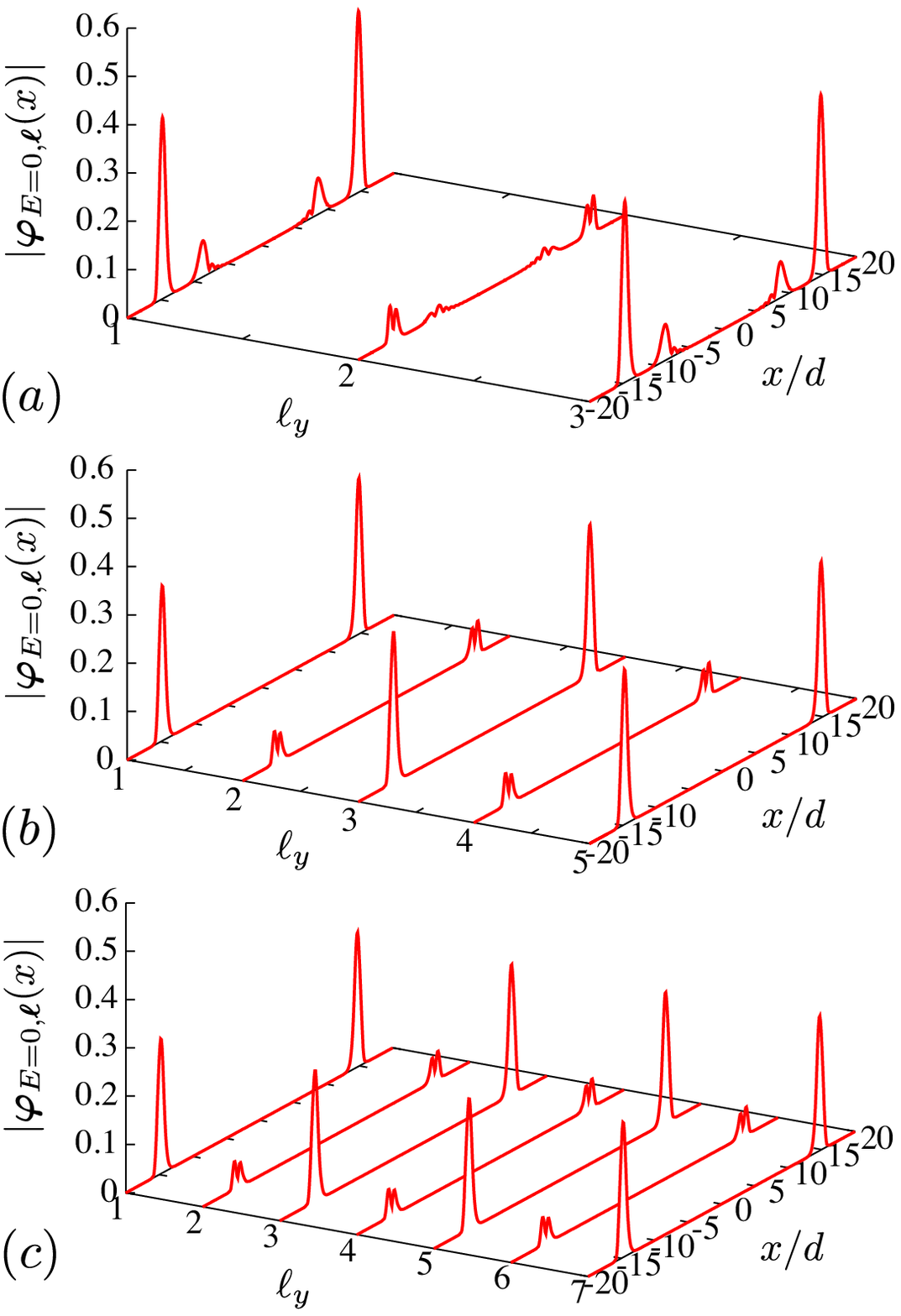}
\caption{Amplitude of wave functions $|{\bm
 \varphi}_{E=0, {\bm \ell}}(x)|$ for the lowest energy state in the case
 of (a)$N_y=3$, (b)$N_y=5$, and (c)$N_y=7$, respectively. 
The parameters are same as
 those in Fig.~\ref{fig:egns}.}
\label{fig:N5_7}
\end{figure}

To understand the low-lying energy states in the case of odd $N_y$'s,
which remain zero modes, we show in Fig.~\ref{fig:ldos} the local density of states (LDOS) at the site ${\bm \ell}$ defined as
\beq
\mathcal{N}_{\sigma, {\bm \ell}}(x,E) &=& \sum _{E_n> 0}
\left[
\left| u_{n,\sigma,{\bm \ell}}(x)\right|^2\delta(E-E_n) +
\left| v_{n,\sigma,{\bm \ell}}(x)\right|^2\delta(E+E_n) 
\right].
\label{eq:ldos}
\eeq
The LDOS for up-spins (down-spins) in the case of $N_y \!=\! 3$ is
displayed in Figs.~\ref{fig:ldos}(a) and (c) ((b) and (d)). It is
clearly seen from Figs.~\ref{fig:ldos}(a) and (b) that the LDOS for
up-spins at ${\ell}_y \!=\! 1$ and $3$ is accompanied by the zero energy states
which are bound at the end points of the tube ($x/d \approx \pm 15$), while
$\mathcal{N}_{\downarrow}$ has sharp peaks at $x/d \!\approx\! \pm 8$
corresponding to the inner phase boundaries between $w(x) \!=\! 0$
(non-topological region) and $1$ (topological region) in the case of
$N_y \!=\! 1$. In contrast, the intertube tunneling split the zero
energy states to the positive and negative energy states in the LDOS at
$\ell_y \!=\! 2$, which have the mini-gap with $\pm 0.02E_{\rm F}$. 
In Fig.\ref{fig:N5_7}, we also show the amplitude of the zero modes
for $N_y=3$, $N_y=5$ and $N_y=7$, respectively. 
In all cases, the wave functions have
large amplitudes at the tubes located at odd $\ell_y$'s.
At even $\ell_y$'s, their amplitudes are almost negligible.

These intertube tunneling effects can be understood as follows.
As was shown above, 
when one neglects the intertube couplings $t_y$, $t_z$ and $\tilde{\kappa}_y$, 
each tube supports four Majorana
zero modes localized at $x\sim \pm 8d$ and $x\sim \pm 15d$.
Now let us denote one of them (say, the zero mode localized at $x\sim \pm 15d$)
as $\gamma_{\ell_y}$ ($\ell_y=1, \cdots, N_y$), and consider how the
intertube couplings affect on them. 
When $t_y$ and $\tilde{\kappa}_y$ are turned on, 
the zero modes on neighboring tubes are coupled by the intertube tunneling, 
\begin{eqnarray}
{\cal
 H}=it\gamma_1\gamma_2+it\gamma_2\gamma_3+\cdots+it\gamma_{N_y-1}\gamma_{N_y},
\label{eq:effective}
\end{eqnarray} 
where $t$ denotes the induced tunneling coupling. 
Note that $t$ is real since $\gamma_{\ell_y}$ is a Majorana zero mode
satisfying $\gamma_{\ell_y}=\gamma_{\ell_y}^{\dagger}$.
Equation (\ref{eq:effective}) is rewritten as ${\cal
H}=\Gamma^{\dagger}\hat{H}\Gamma/2$ with 
\begin{eqnarray}
\hat{H}=
\left(
\begin{array}{ccccc}
0&it&0&\cdots&0\\
-it& 0 &it& \ddots & \vdots \\
0 & -it & 0 & \ddots & 0  \\
\vdots & \ddots & \ddots & \ddots & it  \\
0 & \cdots & 0 & -it & 0
\end{array}
\right)
\end{eqnarray}
and $\Gamma= (\gamma_1, \gamma_2, \cdots, \gamma_{N_y-1}, \gamma_{N_y})^t$. 
Diagonalizing the $N_y\times N_y$ matrix $\hat{H}$, one can examine the
effects of the intertube tunneling.

It can be easily shown that $\hat{H}$ has a single zero
eigenvalue for odd $N_y$'s, while it does not have for even
$N_y$'s.
This result naturally explain why Majorana zero modes survive only for
odd $N_y$'s.
One also finds that the zero eigenstate of $\hat{H}$ has the following form
\begin{eqnarray}
(1,0, -1)^t, \quad \mbox{for $N_y=3$} 
\nonumber\\
(1,0, -1,0, 1)^t, \quad \mbox{for $N_y=5$} 
\nonumber\\
(1,0, -1,0, 1, 0, -1)^t, \quad \mbox{for $N_y=7$}, 
\end{eqnarray} 
which explains qualitatively why the remaining Majorana zero modes
illustrated in Fig.\ref{fig:N5_7} have large amplitudes on tubes at odd
$\ell_y$'s.

The robustness of the zero-energy states against the intertube tunneling
is also understood by the topological number $w$. 
As we mentioned above, even in the presence of intertube tunneling, the
winding number $w$ is well-defined for a whole system of tubes. 
Since one can turn off the intertube tunneling without the bulk gap
closing, the value of $w$ can be evaluated by setting
$t_y=t_z=\tilde{\kappa}_y=0$ in Eq.(\ref{eq:w}). Then one obtains 
\begin{eqnarray}
w={\rm tr}U\frac{1}{2\pi i}\int_{-\infty}^{\infty}dk \partial_k \ln
\left[\det{\cal Q}(k)\right].
\end{eqnarray}
Noting 
\begin{eqnarray}
{\rm tr}U=\left\{
\begin{array}{cc}
0 & \mbox{for even $N_y$'s}\\
N_z & \mbox{for odd $N_y$'s}
\end{array}
\right.,
\end{eqnarray}
one can valuate $w$ as
\begin{eqnarray}
|w|=\left\{ 
\begin{array}{ll}
N_z, & \mbox{for odd $N_y$'s}\\
0, & \mbox{for even $N_y$'s}
\end{array}
\right. ,
\end{eqnarray}
when $\sqrt{h_x^2+h_z^2}>h_{\rm c}$.
This implies that  Majorana zero modes survive for
odd $N_y$'s.

%In the limit of a two-dimensional
%system with $N_y \!\rightarrow\! \infty$, the first Chern number can be
%defined. This Chern number ensures the existence of the Majorana end
%states in each one-dimensional tube. Hence, the consequence of the Chern
%number is inconsistent with the results obtained here that the zero
%energy states survives only at $i \!=\! 1$ and $N_y$ when $N_y$ is odd.   

\section{Conclusions and Discussions}

In this article,  we have investigated the effect of intertube tunneling in a
quasi-one-dimensional Fermi gas with a Rashba-type spin-orbit
coupling. From the argument based on the symmetry of the effective
Hamiltonian, the topological property has been studied.
In the absence of the intertube tunneling, the one-dimensional winding
number in Eq.~(\ref{eq:w}) ensures the existence of Majorana zero modes
bound at the end points of each tube.
It also ensures the existence of the Majorana zero modes in the presence
of the intertube tunneling if the number of tubes in the $y$-direction
is odd.
Using full numerical calculations of self-consistent
equations, we have confirmed that this topological property is clearly
reflected in low-lying quasiparticle states.
% in the case of even
%$N_y$'s. In the case of odd $N_y$'s, however, some of Majorana zero
%modes survive even in the presence of a ${\bm Z}_2$ symmetry breaking
%perturbation. 
%The existence can not be explained by the one-dimensional
%winding number protected by a ${\bm Z}_2$ symmetry and the first Chern
%number, which remains as a future problem. 
These behaviors of low-lying
quasiparticles associated by the ${\bm Z}_2$-symmetry protected topology
might be detectable through the momentum-resolved radio-frequency
spectroscopy~\cite{jin,peng}.  

Here, we have considered the two-dimensional Rashba spin-orbit coupling which has not been realized
in atomic gases yet. It should be mentioned that the results obtained in this work are not straightforwardly applicable to Fermi gases under the realistic spin- orbit coupling with equal Rashba and Dresselhaus strengths~\cite{nist,mit,china}: Because of an additional symmetry specific to the equal Rashba and Dresselhaus spin-orbit coupling, the later situation is accompanied by zero energy states regardless of even-odd parity of Ny. It is also important to mention that topological superfluidity protected by the mirror symmetry can be affected by the orientation of the applied Zeeman field, because the mirror symmetry is explicitly broken. The topological property in a system with the breaking of the mirror symmetry can be associated with the 1D $\mathbb{Z}_2$ number defined in Eq.~(\ref{eq:z2top}). The details will be reported elsewhere~\cite{TMfuture}.

Finally, we would like to mention about a generalization of the present
consideration to  semiconductor-superconductor nanowire systems.
The one-dimensional winding number (\ref{eq:w}) introduced in this
paper is also applicable to semiconductor-superconductor nanowire with
multichannels. 
If we consider the nanowire extending in the $x$-direction on the top of
an $s$-wave superconductor in the $xy$-plane, the system is naturally
supposed to be invariant
under the mirror reflection, $y\rightarrow -y$, to the $xz$-plane.
This mirror symmetry could be broken under Zeeman fields, but the ${\bm Z}_2$
symmetry (\ref{eq:z2}) remains if the Zeeman fields are applied in the $x$- or
$z$-direction. 
Then, the topological number $w$ in Eq. (\ref{eq:w}) is defined in the
same manner.
From arguments similar to the above, one finds that $w$ is nonzero
if the Zeeman field $h$ satisfies $|h|>h_{\rm c}$ and 
the number of the channels in the $y$-direction of the nanowire is odd.
Indeed, under this condition, $|w|$ is equal to the number
of channels in the $z$-direction of the nanowire. 
Note that, in contrast to the 1D ${\mathbb Z}_2$ number in Eq.(\ref{eq:z2top}), $w$ 
can be nonzero even when the total number of channels in the nanowire is
even, since it is given by sum of the channels in the $z$ and $y$-directions.
As well as the fermionic gas system studied in this paper, the local
density operator of the Majorana zero modes vanishes \cite{TM2012}.
This implies that the coupling between the Majorana zero modes and
non-magnetic local disorder potential also vanishes, and thus
the Majorana zero are stable against weak non-magnetic
disorders.

% found to be analogous to a
%
%semiconductor-superconductor nanowire with multibands. 
% However, the intertube coupling through $t_y$ and
%$\tilde{\kappa}_y$ explicitly breaks the ${\bm Z}_2$, where $w$ is
%ill-defined. 

\section*{Acknowledgments}
The authors are grateful to Takuto Kawakami for fruitful discussions and
comments.
This work was supported by a Grant-in-Aid for Scientific Research from MEXT
of Japan, ``Topological Quantum Phenomena'' No.~22103005 and JPSJ
No.~22540383.

\bibliographystyle{model1a-num-names}
%\bibliography{ebi}



\end{document}